\documentclass[aps,prd,nofootinbib,twocolumn,superscriptaddress,preprintnumbers,balancelastpage,longbibliography]{revtex4-1}

\usepackage{amsmath,amssymb}
\usepackage{graphicx}
\usepackage{color}
\usepackage{units}
\usepackage[hyperfootnotes=false,colorlinks,citecolor=blue]{hyperref}
\usepackage{soul}
\usepackage[normalem]{ulem}

\newcommand{\beq}{\begin{equation}}
\newcommand{\eeq}{\end{equation}}
\newcommand{\bea}{\begin{eqnarray}}
\newcommand{\ena}{\end{eqnarray}}

\newcommand{\Msun}{{\ifmmode{{\rm{M_{\odot}}}}\else{${\rm{M_{\odot}}}$}\fi}} 

\allowdisplaybreaks

\begin{document}

\title{Faint light of old neutron stars and detectability at the James Webb Space Telescope}

\author{Shiuli Chatterjee}
\email{shiulic@iisc.ac.in}
\affiliation{Centre for High Energy Physics, Indian Institute of Science, Bangalore 560012, India}
\author{Raghuveer Garani}
\email{garani@fi.infn.it}
\affiliation{INFN Sezione di Firenze, Via G. Sansone 1, I-50019 Sesto Fiorentino, Italy}
\author{Rajeev Kumar Jain}
\email{rkjain@iisc.ac.in}
\affiliation{Department of Physics, Indian Institute of Science, Bangalore~560012, India}
\author{Brijesh Kanodia}
\email{brijeshk@iisc.ac.in}
\affiliation{Centre for High Energy Physics, Indian Institute of Science, Bangalore 560012, India}
\affiliation{Department of Physics, Indian Institute of Science, Bangalore~560012, India}
\author{M. S. N. Kumar}
\email{nanda@astro.up.pt}
\affiliation{Instituto de Astrofísica e Ciências do Espaço, Porto, Rua  das  Estrelas,  s/n,  4150-762,  Porto, Portugal}
\author{Sudhir K. Vempati}
\email{vempati@iisc.ac.in}
\affiliation{Centre for High Energy Physics, Indian Institute of Science, Bangalore 560012, India}

\begin{abstract}
Isolated ideal neutron stars (NS) of age $>10^9$ yrs exhaust thermal and rotational energies and cool down to temperatures below $\mathcal{O}(100)$ K.  Accretion of particle dark matter (DM) by such NS can heat them up through kinetic and annihilation processes. This increases the NS surface temperature to a maximum of $\sim 2550$ K in the best case scenario.  
The maximum accretion rate depends on the DM ambient density and velocity  dispersion, and on the NS equation of state and their velocity distributions. 
Upon scanning over these variables, we find that the effective surface temperature varies at most by $\sim 40\%$. Black body spectrum of such warm NS peak at near infrared wavelengths with magnitudes in the range potentially detectable by the James Webb Space Telescope (JWST). Using the JWST exposure time calculator, we demonstrate that NS with surface temperatures $\gtrsim 2400$ K, located at a distance of 10\,pc can be detected through the F150W2 (F322W2) filters of the NIRCAM instrument at SNR\,$\gtrsim 10$ (5) within 24 hours of exposure time. Independently of DM, an observation of NS with surface temperatures $\gtrsim 2500$ K will be a formative step towards testing the minimal cooling paradigm during late evolutionary stages.
\end{abstract}

\maketitle

\paragraph*{{\it 1. Introduction.---}} Neutron stars (NS) are one of the most compact and dense astrophysical objects in the universe. They are stellar configurations that are supported against gravitational collapse by the neutron degeneracy pressure and that from neutron-neutron interactions ~\cite{Shapiro:1983du}. NS are notoriously hard objects to detect observationally, especially if they are very old. Most of the detected NS  $\sim \mathcal{O}$(10$^3$) have been observed as isolated radio pulsars  with ages $\lesssim$ 100 Myrs. Older NS, have been detected so far, only as companions in binary systems, e.g. millisecond pulsars~\cite{Lorimer:2008se,yakovlev2011cooling,Lattimer:2015nhk}. A better understanding of these objects is emerging, thanks to recent gravitational wave observations of binary NS mergers~\cite{LIGOScientific:2018cki}, and other studies estimating their mass and radius~\cite{riley2019nicer,miller2019psr,riley2021nicer,miller2021radius}. Old ($>10^9$ yrs) isolated  NS have not been identified so far. According to the minimal cooling paradigm~\cite{Page:2005fq} such objects exhaust all their thermal and rotational energy, making it impossible to observe them.
A key feature of passive cooling models or minimal cooling paradigm is that they predict a drastic fall in the temperature once the neutrino emission from the core becomes smaller than that of thermal photons from the surface. This leads to surface temperatures below $10^4$ ($10^3$) K after about $10$ (100) million years, see ref.~\cite{Page:2005fq} and references within. 
As a consequence, very little is known about their physical and statistical properties~\cite{blaes1993ApJ,Treves:1999ne,Ofek:2009wt,2010A&A...510A..23S,2010A&Agonzalesheat,Tani2016RAA}.

The situation can change when dark matter (DM) particles in the galactic halo are efficiently captured by NS. These DM particles can deposit energy via scattering and annihilation processes, giving rise to kinetic~\cite{Baryakhtar:2017dbj} and annihilation heating~\cite{Kouvaris:2007ay,2010A&Agonzalesheat} of the NS, respectively. The resulting increase in the surface temperature of NS can be significantly larger than the case without DM heating. In the local bubble, NS older than Gyrs can have an increase in temperature  from about $\sim 100$ K to $\sim 2000$ K due to DM heating mechanisms~\cite{Baryakhtar:2017dbj,Bell:2018pkk}. Observation of such NS would place constraints on DM mass and their interaction strength with visible matter in a broadly model independent way~\cite{Kouvaris:2007ay,Kouvaris:2010vv,Baryakhtar:2017dbj,Garani:2018kkd,Bell:2018pkk,Garani:2019fpa,Bell:2019pyc,Bell:2020jou,Bell:2020lmm,Raj:2017wrv,Acevedo:2019agu,Bell:2020obw,Anzuini:2021lnv,Tinyakov:2021lnt,Fujiwara:2022uiq}.   

During late evolutionary stages ($t_{\rm age}>$ Gyrs), the luminosities of NS that accrete DM are solely controlled by the DM capture rate.
This depends on several factors including the physics of the nuclear matter at near saturation densities, NS and DM phase space distributions, and the interaction strength of DM with visible matter, and their masses. For any given neutron star of mass M and radius R, the maximal rate of DM accretion is given by the so-called geometric rate corresponding to scattering cross section $\sigma^{\rm g}_\star=\pi R^2 /N \sim 10^{-45}$ cm$^2$, where $N$ is the total number of target neutrons in the NS.\footnote{The value of the geometric cross section is mildly DM model dependent, which can vary by up to two orders of magnitude~\cite{Bell:2020obw,Anzuini:2021lnv}. }

In this {\it letter}, we evaluate the DM capture rate (and consequent change in the effective surface temperature of NS) for a range of equations of state (EoS) allowed by current observational data, and a viable range of NS and DM phase space distributions. We find that the maximal temperature is $\sim 2550$ K and varies only by 40\% over this whole range of inputs. Subsequently, we study the prospect of observing such old maximally heated NS close to the Solar position. The spatial distribution of such objects are predicted using Monte-Carlo orbital simulations of galactic  NS~\cite{blaes1993ApJ,Treves:1999ne,Ofek:2009wt,2010A&Agonzalesheat,2010A&A...510A..23S,Tani2016RAA}. These simulations suggests that we can expect $1-2$ ($100-200$) old isolated NS within 10 (50) pc. The black body spectrum of NS that are maximally heated by DM peaks at near infrared wavelengths of $\lambda \sim 2 \mu\,m$, and fall in the range of observability for the recently launched James Webb Space Telescope (JWST)~\cite{Baryakhtar:2017dbj}. Using the publicly available JWST exposure time calculator~\cite{TMTInternationalScienceDevelopmentTeamsTMTScienceAdvisoryCommittee:2015pvw}, we discuss the optimal observation strategy that can be employed to hunt such old NS. Among the widest band filters available onboard JWST, we find that the NIRCAM filter F150W2 provides the best sensitivity and facilitates detection  with signal to noise ratio (SNR) $\gtrsim 10$ within 24 hours of exposure time.  \\

\paragraph*{{\it 2.  Maximal DM capture in NS.---}} The maximum rate at which DM particles can be gravitationally captured by the NS depends on its mass ($M$), radius ($R$), velocity ($v_\star$), the DM dispersion velocity ($v_d$) and the ambient DM density ($\rho_\chi$).  
The geometric limit is considered for the capture rate when the DM mean free path is approximately smaller than the neutron star radius, which corresponds to DM scattering cross section $\sigma \gtrsim \sigma^{\rm g}_\star$ ~\cite{Goldman:1989nd, Kouvaris:2007ay,Bottino:2002pd, Bernal:2012qh, Garani:2017jcj, Bell:2018pkk,Tinyakov:2021lnt}. 
For a given DM mass ($m_\chi$), the maximum capture rate  (geometric rate)
is given by~\cite{Bell:2019pyc}
\begin{equation}
\label{eq:capturegeom1}
C_\star^{\rm g}=\pi R^2 \, \frac{\rho_\chi}{m_\chi}\, \frac{\langle v \rangle_0}{1-v^2_{esc}}\sqrt{\frac{3 \pi}{8}} \frac{v^2_{esc}}{v_\star v_d} \,{\rm Erf}\left(\sqrt{\frac{3}{2}} \frac{v_\star}{v_d}\right),
\end{equation}
with $\langle v \rangle_0 = \sqrt{8/(3\pi)}v_d \,$, the escape velocity $v_{esc}=\sqrt{2GM/R}$ from the neutron star, and the error function is denoted by ${\rm Erf}$. Thus the accretion rate strongly depends on the EoS of the neutron star through $v_{esc}$, and modestly on variables $\rho_\chi$, $v_\star$ and $v_d$. When the DM mean free path is approximately larger than the NS radius, the geometric capture rate above should be rescaled by factor $\sigma/\sigma^{\rm g}_\star$. In the opposite limit, when the mean free path is smaller than the NS radius and $m_\chi>10^6$ GeV, DM requires more than one scattering to lose enough energy to be captured. Taking this into account intrinsically changes the $m_\chi$ dependence in eq.~\eqref{eq:capturegeom1}~\cite{Bramante:2017xlb,Baryakhtar:2017dbj}. The main goal of this work is to systematically evaluate how the maximum capture rate in the single scattering regime, and consequently, the maximal neutron star heating induced by DM capture and annihilation, are affected by the allowed range of these parameters. \\

\paragraph*{{\it 2.1 Maximal DM heating of NS:}} Ambient DM in the halo is accelerated to semi-relativistic velocities as it impinges on the NS. Scattering of the DM particles with particles in the stellar medium ($n,p,\mu,e$)  can result in the capture of DM in NS as most of their initial kinetic energy is deposited to NS medium, thereby heating it up~\cite{Baryakhtar:2017dbj}. Next, if DM particles can annihilate during the current cosmological epoch, the accumulated DM in the core of NS would also annihilate upon thermalization with NS medium\footnote{For geometric values of cross section, DM particles thermalize within $\mathcal{O}(10)$ Myrs for $10^{-8}\, {\rm GeV} < m_\chi< 10^4\, {\rm GeV}$~\cite{Bertoni:2013bsa,Acevedo:2019agu,Garani:2021feo}. Capture and annihilation processes are in equilibrium as long as s-wave (p-wave) annihilation cross sections are greater than $\sim 10^{-53}$ cm$^3$/s ($10^{-44}$ cm$^3$/s) for Gyr old NS.  }~\cite{Kouvaris:2007ay,Kouvaris:2010vv,Bertoni:2013bsa,Garani:2020wge}. 
Further heating of the NS is possible, if the products of annihilation process deposit all their energy in the NS medium. In this case, the total energy deposited is equal to the sum of kinetic and annihilation energies, dubbed here as KA heating. The kinetic and annihilation energies deposited are $m_\chi (\gamma -1) C^{\rm g}_\star$ and $m_\chi C^{\rm g}_\star$, respectively.  Here, $\gamma = (1- 2GM/R)^{-1/2}$ is the gravitational redshift factor. Therefore, the deposited energies are independent of DM mass. Note, however, there exists a lower limit on the DM mass below which DM evaporation from the NS dominates~\cite{Bell:2013xk,Garani:2018kkd,Bell:2020jou,Garani:2021feo}. This lower limit is set when the ratio of escape energy from the core to the core temperature is $\sim30$~\cite{Garani:2021feo}.  For isolated and cold NS,  the kinetic only, or the KA energy can effectively render the NS observable by increasing the surface temperature which translates directly to the luminosity, as given by the Stefan-Boltzmann law ($4 \pi R^2 \sigma_{\rm B} T_{\rm eff}^4$). For an observer far away from the NS, the apparent temperature is $T^\infty=T_{\rm eff}/\gamma$. The contributions from kinetic only, and KA heating can be computed as
\begin{eqnarray}\label{eq:teffk}
T_{\rm kin}^{\infty} &\approx&  1787\,\rm{K}\Bigg[\frac{\alpha_{\rm kin}}{0.08} \left(\frac{\rho_\chi}{0.42 \,\rm{GeV/cm^3}}\right) \Bigg. \\
&& \hspace{-0.7cm}\Bigg. \times \left(\frac{220\,\rm{km/s}}{v_\star}\right) {\rm{Erf}}\left( \frac{270\,\rm{km/s}}{v_d} \frac{v_\star}{220\,\rm{km/s}} \right) \Bigg]^{1/4}\, , \nonumber
\end{eqnarray}
\begin{eqnarray}\label{eq:teff1}
T_{\rm KA}^{\infty} &\approx& 2518~\rm{K}\Bigg[\frac{\alpha_{\rm KA}}{0.33} \left(\frac{\rho_\chi}{0.42 \,\rm{GeV/cm^3}}\right)  \Bigg. \\ 
& & \hspace{-0.7cm}\times \Bigg. 
\left(\frac{220\,\rm{km/s}}{v_\star}\right) {\rm{Erf}}\left( \frac{270\,\rm{km/s}}{v_d} \frac{v_\star}{220\,\rm{km/s}} \right) \Bigg]^{1/4}\, , \nonumber 
\end{eqnarray}
with,
\beq\label{eq:alpha}
\alpha_{\rm kin} =  \frac{(\gamma -1)(\gamma^2-1)}{\gamma^4} \quad \quad \rm{and} \quad \quad \alpha_{\rm KA} = \frac{\gamma (\gamma^2-1)}{\gamma^4}~.\nonumber
\eeq
 The above expressions are normalized to $M= 1.5 \,M_\odot$ and $R = 10$ km. Note that $\alpha_{\rm KA}$ ($\alpha_{\rm kin}$) is maximized for $\gamma= 1.732\,(2.56)$.  If old NS exist in binary systems, then, depending on the period of their orbits, the DM capture rates~\eqref{eq:capturegeom1} will be enhanced by factors up to 4~\cite{Brayeur:2011yw}.

Mechanisms of kinetic and KA heating can heat the NS up to $\mathcal{O}$($10^3$) K after about 100\,Myrs. However, for similar ages, other mechanisms of heating can begin to work, 
if exotic phases such as neutron/proton superfluidity are realized in their cores. Generically, heat can be injected into the NS by conversion of magnetic, rotational, and/or chemical energies~\cite{Page:2005fq,2010A&Agonzalesheat}. 
In such scenarios, the intrinsic heating signature is degenerate with that of DM heating, challenging its interpretation. 
Constraints on DM parameter space can nevertheless be placed if NS with effective surface temperature $\lesssim 2 \times 10^3$ K were to be observed~\cite{Hamaguchi:2019oev,Yanagi:2020yvg}. \\

\paragraph*{{\it 3. Computational Inputs.---}}\label{sec3} 
In this section we will briefly discuss the inputs necessary to compute the quantities described by eqs.~\eqref{eq:capturegeom1},\eqref{eq:teffk}, \& \eqref{eq:teff1}. The primary variables are  $v_\star$, $\rho_\chi$ and $v_d$, and the NS EoS through $\gamma$.  We detail the sources from where the range of values are obtained together with observational and or theoretical justifications in the supplementary material.

We consider values of $\rho_\chi$ and $v_d$ obtained from existing fits to dynamical measurements of  rotational curves, inclusive of baryonic effects, assuming generalized NFW profile, and Maxwell-Boltzmann velocity distribution in the galactic frame~\cite{Pato:2015dua}. We choose values of $\rho_\chi= 0.39 - 0.52$ GeV/cm$^3$ and corresponding values of $v_d$  in range $260 - 316$ km/s.   

The expected velocities of old NS in the local bubble are estimated from population synthesis studies of such objects in the Milky way\,\cite{blaes1993ApJ,Ofek:2009wt,2010A&A...510A..23S, Tani2016RAA}. We consider two normalized probability distribution functions (PDFs) obtained in ref.~\cite{Ofek:2009wt}  (models unimodal and bimodal) and three from ref.~\cite{2010A&A...510A..23S} (models 1C, 1E and 1E$^\star$), which are representative of the full range of possibilities. The median values of the distribution functions are $214.9, 196.2, 179.18, 202.3$, and  $233.6$ km/s, respectively. 

To compute the gravitational redshift factor, $\gamma = (1- 2GM/R)^{-1/2}$,  it is necessary to consider the full range of mass and radius allowed by viable NS EoS. 
A wide variety of NS EoS, taken in the limit of zero temperature, are studied and discussed in the literature~\cite{Cohen1970,Page:2004fy,Goriely:2010bm,Goriely:2013xba,Potekhin:2013qqa,Drischler:2016cpy,Lattimer_2001}. In this work we consider Mass-Radius relationship coming from representative EoS WFF-1, BSK-21, AP3, AP4, MPA-1, PAL-1 and H4, which are allowed by current astrophysical and gravitational wave data. The values of M-R span the range $0.5-2.2\,M_\odot$ and $9.4-14.2$ km, respectively.\footnote{Note that population synthesis models suggest occurrence of heavier NS ($>2\,M_\odot$) is rarer than lighter ones.} \\

\begin{figure*}[t] 
	\includegraphics[height=0.45\textwidth]{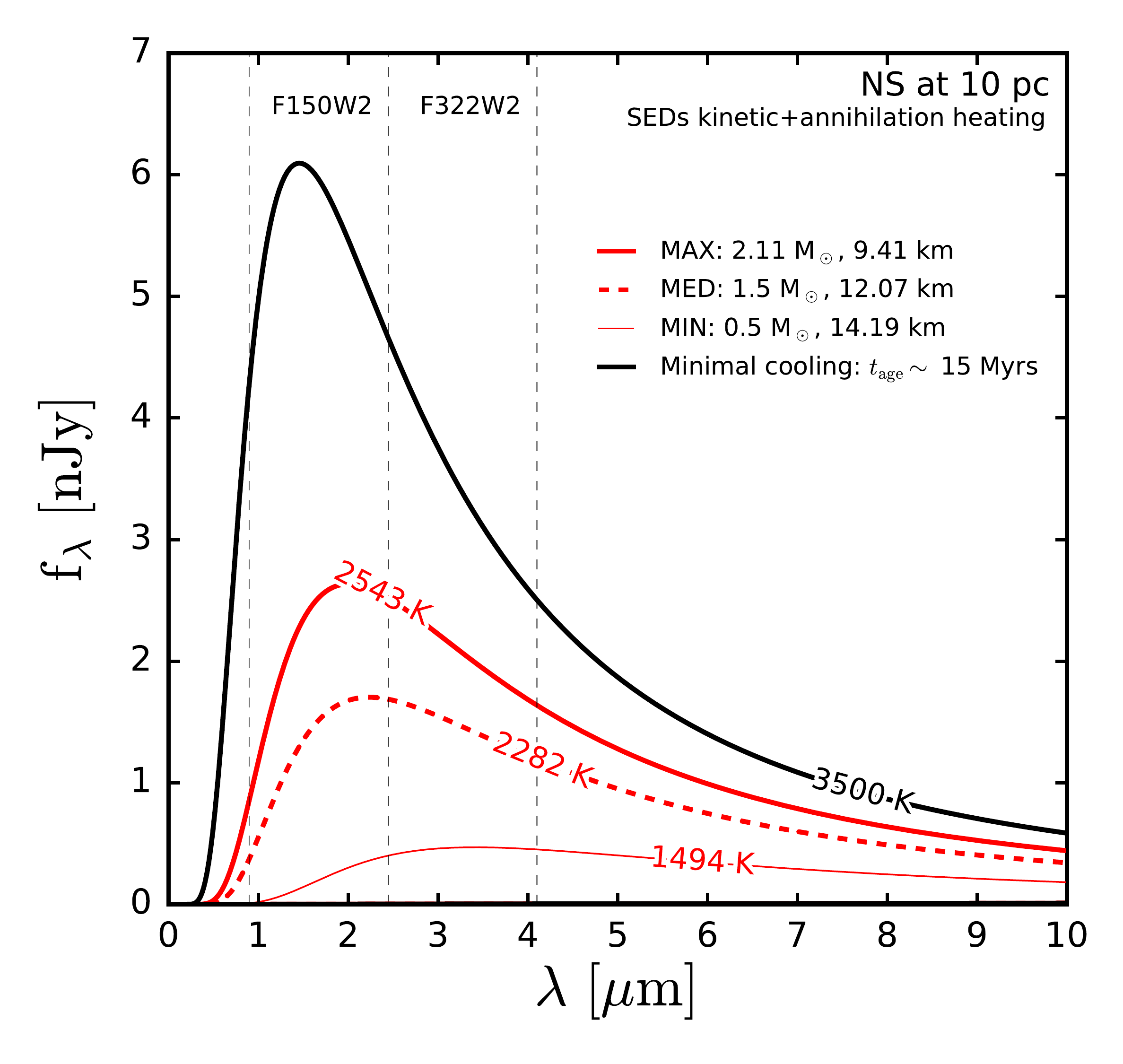}   
	\includegraphics[height=0.45\textwidth]{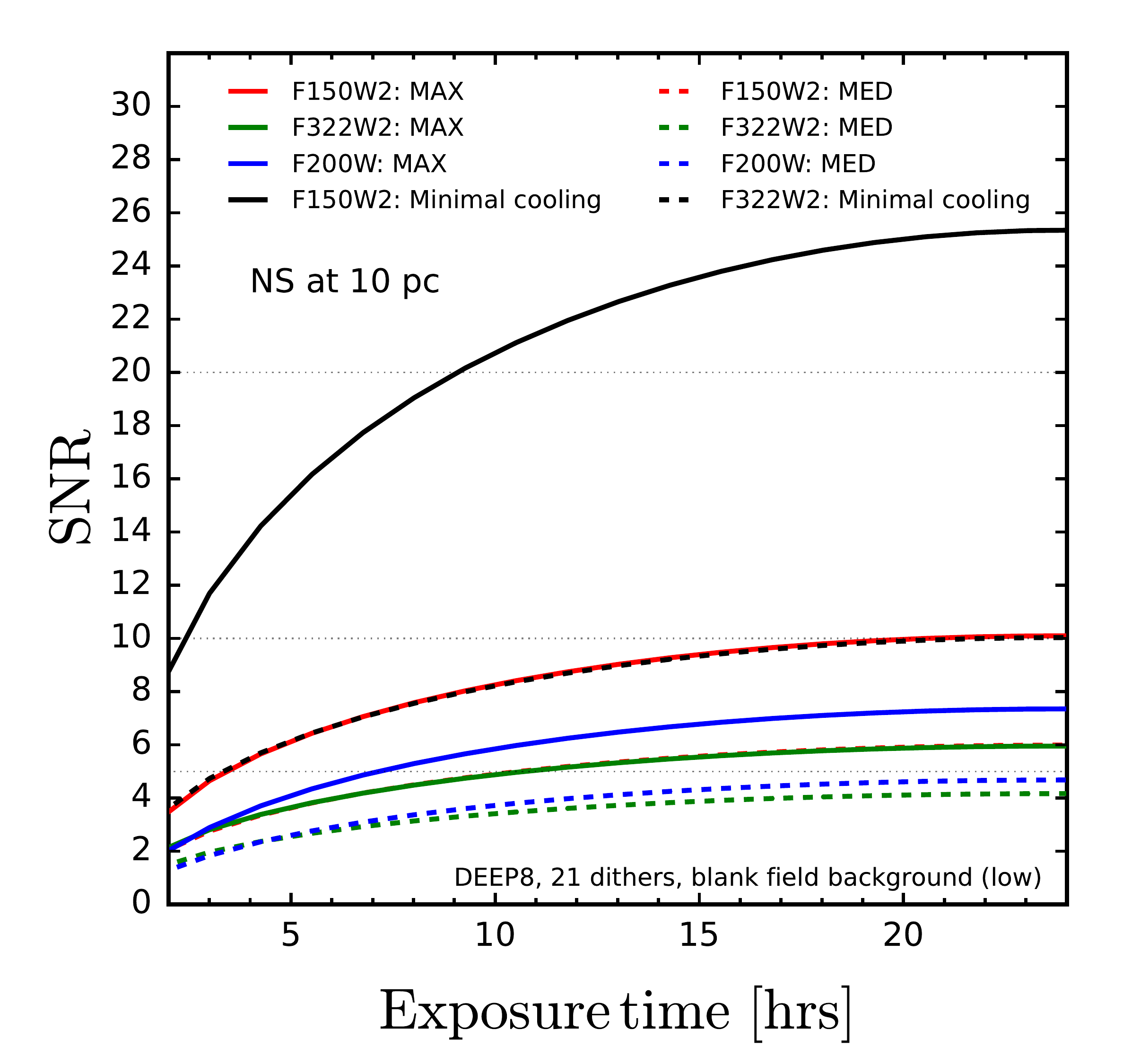} 
	\caption{{\it Left panel:} Range of black body spectral energy distributions for the case of kinetic+annihilation heating, for old isolated NS at 10 pc are shown in red. The thick black line is representative of minimal cooling model for NS of age $\sim 15$ Myrs with black-body temperature $3500$ K, with mass (radius) 1.5 $M_\odot$ (12 km). The thick solid line, dashed line, thin line, correspond to effective temperatures of  2543 K, 2282 K, 1494 K, and mass (radius) of $M=2.11 \, M_\odot$ (9.41 km), $M=1.5\,M_\odot$ (12.07 km), $M=0.5\,M_\odot$ (14.19 km), respectively. Dashed lines were obtained upon averaging over all EoS independent inputs, while the solid (thin) line is representative of the maximum (minimum) value of effective temperature over all EoS we consider over the mass range $0.5-2.2$ $M_\odot$. Vertical dashed lines delimit the bandwidth of filters F150W2 and F322W2. {\it Right panel:} The signal-to-noise ratio (SNR) is shown as a function of exposure time for filters F150W2 and F322W2, and narrow band filter F200W. See text for details.  }
	\label{fig:binary+ka_f}
\end{figure*}

\paragraph*{\it 4. Luminosities of DM accreting NS.---} Using the computational inputs discussed in section 3, we now compute the luminosities of NS from DM accretion heating. For each set of (M, R) given by EoS $i$ we compute the DM geometric capture rate averaging over DM phase space parameters through
\begin{align}\label{eq:gcapavg}
C^{\rm g}_{i,j}(M, R) = \kappa \sum_{k,l} \int d v_\star p_j(v_\star) C^{\rm g}_\star(i,v_\star, v^k_d,\rho_\chi^l)~,
\end{align}
where, $j=1,5$ corresponds to the different velocity probability distributions ($p_j$) of NS discussed above, and the averaging coefficient $\kappa = (k_{\rm max} \,l_{\rm max})^{-1}$. Integers $k_{\rm max}$(=2) and $l_{\rm max}$(=2) denote the number of values we sample for parameters $v_d$ and $\rho_\chi$, respectively. When both kinetic and annihilation heating processes are operative, the effective surface temperature $T^\infty_{i,j}$ is obtained by summing both contributions $(m_\chi(\gamma-1) + m_\chi )C^g_{i,j}$ and equating it to the apparent luminosity (see eq.~\eqref{eq:teff1}). Next, we average over the NS velocity distributions by $\langle C^g_{i}\rangle = \sum_jC^g_{i,j}/j_{\rm max}$ to get an  effective average surface temperature $T^\infty_{{\rm avg},i}$  . 
Assuming the NS to be a black body, we compute the spectral energy distribution (SED) as follows
\begin{equation}\label{eq:sed}
f_\lambda(M,R) = \frac{4 \pi^2}{\lambda^3}\left(e^{\frac{2 \pi}{\lambda T^\infty}} -1\right)^{-1} \left(\frac{R\, \gamma}{d}\right)^2~.
\end{equation}
Here $R$ and $d$ are the radius (typically 10\,km) and distance (taken to be $d$=10\,pc) to the NS, and the factor $R \gamma /d$ is the angle subtended by the NS to the observer. The flux density at a given wavelength $\lambda$ is denoted by $f_\lambda(M,R)$, and $T^\infty$ is the surface temperature of the NS (see eq.~\eqref{eq:teff1}). 
In Fig.~\ref{fig:binary+ka_f} (left panel), we display the SEDs due to KA heating, encompassing the temperature range for each combination of mass and radius shown in the right panel of Fig.~\ref{fig:EoS_M_R_all}, and NS velocity PDFs discussed above, also shown in the left panel of Fig.~\ref{fig:EoS_M_R_all}. We also present the SED (black) for a NS of mass (radius) 1.5 $M_\odot$ (12 km), for AP4 EoS, with age $\sim 15$ Myr, representative of late evolutionary stages within the minimal cooling paradigm~\cite{Page:2005fq}.

We consider three scenarios for the NS luminosity.  The MAX scenario is obtained for the NS EoS WFF-1 and the 1E velocity PDF, while, the MIN scenario is realized by EoS PAL-1 and bimodal velocity PDF. The MED scenario, however, is obtained by averaging over all NS velocity PDFs, and corresponds to EoS AP3. The maximal difference in the temperature between the MAX and MIN scenario is $\sim 40\%$.

\paragraph*{{\it 4.1 Detectability through JWST:}} As evident from Fig.~\ref{fig:binary+ka_f} (left panel), the SEDs peak at $\lambda\sim$\,2\,$\mu$m in the near-infrared bands with maximum flux values $\sim 2.5 \,{\rm nJy}$. Owing to the compact ($R \sim$ 10\,km) size of NS, they will appear as unresolved, extremely faint point sources, even for a cutting-edge facility such as the JWST. 
Any attempts to detect such objects warrant exploiting the full potential of JWST, optimizing every available resource within the telescope and cameras.
In the right panel of Fig.~\ref{fig:binary+ka_f}, we demonstrate the sensitivity of the Near-Infrared Camera (NIRCAM) on the JWST, to the MAX, MED, and minimal cooling SEDs shown in the left panel.
The signal-to-noise ratio (SNR) is plotted as a function of the exposure time, where the SNR is computed using the exposure time calculator (ETC) specifically dedicated for JWST and WFIRST missions~\cite{2016SPIE.9910E..16P}. 
The SEDs displayed in Fig.~\ref{fig:binary+ka_f} (left) are injected to the ETC as source flux distributions. With the use of JWST background tools we generate a reference (low) background model for a blank field given by coordinates RA='03 32 42.397'  and   Dec= '-27 42 7.93'. NIRCAM offers only two very wide-band filters, that allow for effective collection of large amounts of photons. They are F150W2 and F322W2, roughly corresponding to the near-infrared H and L bands, and their wavelength coverage is marked in Fig.~\ref{fig:binary+ka_f} (left) using vertical dashed lines.
The very faint nature of the targets require integration times longer than 1000\,s, therefore observations will require the DEEP8 readout pattern\footnote{NIRCAM allows for 9 different data readout patterns. The DEEP8 pattern involves taking an image with 20 samples per group where 8 frames are averaged and 12 frames are skipped, in each group. Maximum number of groups for this mode is 20. Each detector readout takes 10.737 s for the full frame of 2048$\times$2048 pixels~\cite{2016jdox.rept}.} in order to minimize data volume. Given that this readout pattern will be strongly affected by cosmic-rays, we assumed 21 dithers\footnote{Dithering is a technique in which multiple images are obtained with small projected angular offsets on the sky. The resulting images are then (median) combined to eliminate detector artifacts~\cite{2016jdox.rept}, and the spurious cosmic-ray hits, that would affect the individual images.}.
From this figure, it is seen that NS with maximal KA heating (MAX) has excellent prospects to be detected at SNR $\sim$ 5--10 in $<$24\,hrs of observing time with F150W2. Detecting MED scenario is possible only at a SNR$\sim 5$ after 24 hrs of exposure (25hr limit for a JWST small program) for the filter F150W2. Assuming a reference (high) background reduces the maximum attainable SNR within a day of observation by at most two units, see section S2 in the supplementary material for further details.\\

\begin{figure}[h]
    \centering
   \hspace{-1.3cm} 
   \includegraphics[scale=0.35]{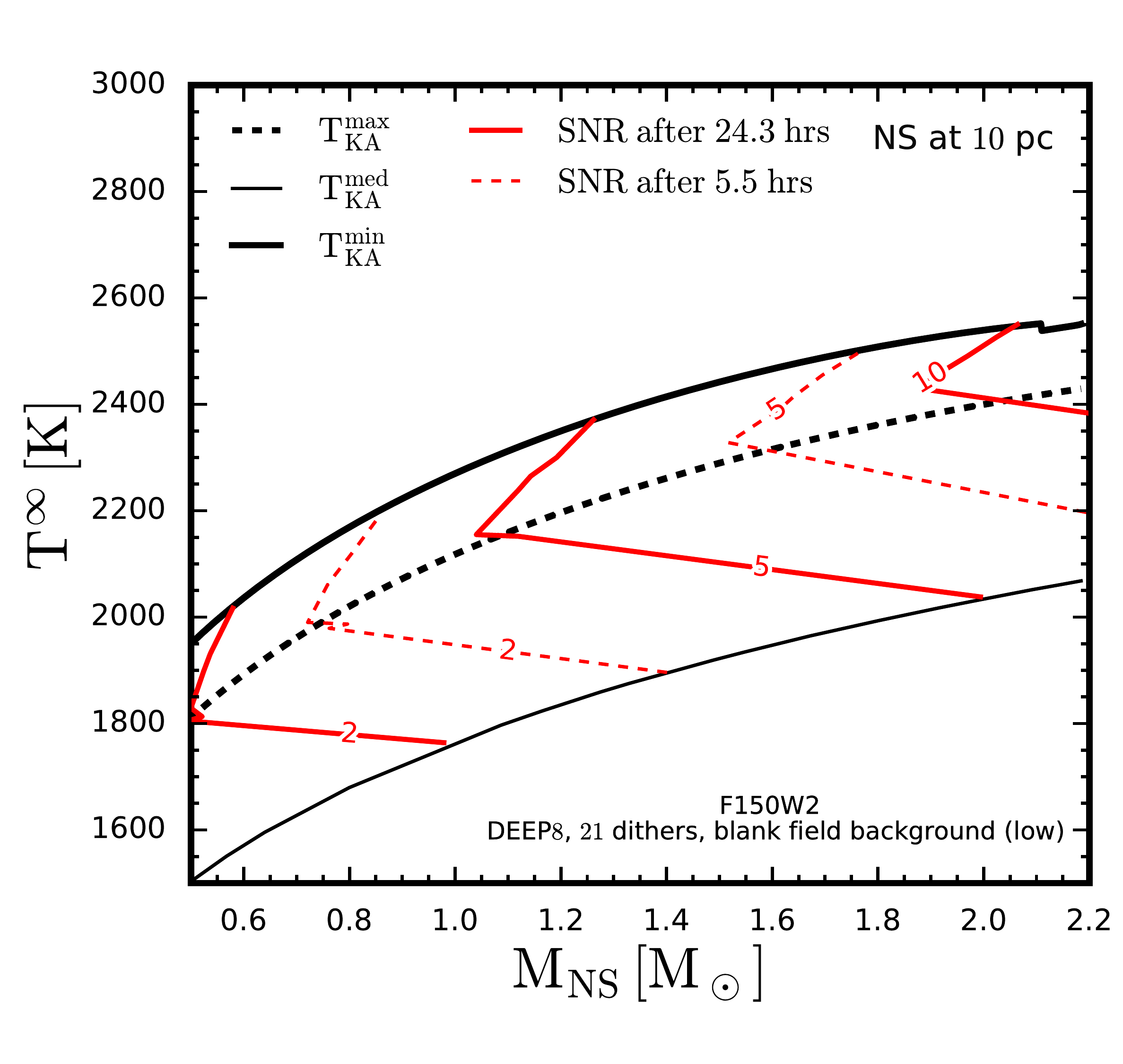}
    \caption{The effective temperature due to kinetic+annihilation heating, for old isolated NS at 10 pc, is shown as a function of its mass. The black solid thick (thin) curves are the maximum (minimum) possible temperature obtained upon averaging over parameters $v_d$ and $\rho_\chi$. The black dashed line denotes the effective  temperature obtained after further averaging over NS velocity PDFs. The contours of SNR corresponding to exposure time of 24.3 (5.5) hrs, for the filter F150W2 are shown in red solid (dashed) lines. The readout mode, reference background model and the number of dithers are the same as in Fig.~\ref{fig:binary+ka_f}. }
    \label{fig:res_geom}
\end{figure}

\paragraph*{{\it 4.2 Observational prospects:}} In Fig.~\ref{fig:res_geom}, we summarize the prospects of detecting KA heated isolated old NS located at 10\,pc distance. In the effective temperature vs NS mass plane, we plot contours of SNR = 2,\,5,\,10 for exposure times of 24.3 hrs (5.5 hrs) in solid (dashed) red, obtained by imaging through the F150W2 filter. The absolute maximum (minimum) temperatures for each NS mass is plotted with black thick (thin) curves. The dashed line represents the temperature obtainable upon averaging over the inputs $\rho_\chi$, $v_d$, and $v_\star$ for EoS AP3. We find that the prospect for detection at SNR $\gtrsim10$ of KA heated isolated NS is realized for the heaviest NS $\sim 2\, M_\odot$ in our sample, corresponding to $T^\infty \gtrsim 2400$ K.  While the prospect for detection of lightest of the NS $\lesssim 1 \, M_\odot$  in our sample is not encouraging. 

For a given M-R, the SNR typically increases with the surface temperature, however, close to the $T^{\rm max}_{\rm KA}$ curve the SNR contours display features that bend rightwards. This is because the luminosity is proportional to the radius of the NS which varies while scanning through the EoS. The kink in $T^{\rm max}_{\rm KA}$ curve at $M_{\rm NS}=2.1\,M_\odot$ is due to a jump from the end point of EoS WFF-1 to another.

For the scenario involving only kinetic heating, the maximum temperature is $\sim 2045$ K for a NS of $2.1\,M_\odot$ and 9.41 km, corresponding to EoS WFF-1 and NS velocity PDF 1E. For this case, after 24.3\,hrs of exposure through the F150W2 filter, a maximum SNR $\sim 4.5$ can be obtained. For the minimal cooling scenario, corresponding to relatively young NS of age $15 $ Myrs, with black-body temperature $3500$ K, SNR of $10$ can be reached within $\lesssim 4$ hrs of observation. As shown in this section, detecting old and cold NS with JWST has good prospects, however, observational implementation will require assembling candidate target lists for such objects through deep surveys from space and ground based facilities, such as the WFIRST or Vera C.\,Rubin observatory. \\

\paragraph*{\it 5. Conclusion.---} DM from the halo can accrete on to old NS leading to anomalous heating of these objects during late evolutionary stages~\cite{Kouvaris:2007ay}. Leveraging this argument, it was pointed out in ref.~\cite{Baryakhtar:2017dbj} that NS heated by DM could be observable by a state-of-the-art facility such as the JWST, and that the discovery of sufficiently cold and old NS in the local bubble would constrain the interactions of particle DM with the Standard Model in a model independent way.

In this {\it letter}, we have {\it quantified } the observational prospects for maximal (kinetic and annihilation) DM heating scenario of NS, corresponding to geometric values of scattering cross section $\sim 10^{-45}$ cm$^2$, and for the minimal cooling scenario representative of NS of age $\sim 15$ Myrs. The effective surface temperature of NS due to DM heating depends on the NS mass and radius through the EoS, the NS velocity in the local bubble, DM velocity and number density. The corresponding variations of the NS effective surface temperature have been assessed together with its impact on the observability of such NS using the NIRCAM instrument on the JWST. Our study implies that NS  warmer than $\gtrsim$ 2600 K, in the local bubble, are observable with strategies requiring shorter exposure times than discussed here. Such scenarios can be realized in a particle physics model dependent way through `Auger effect', if DM is charged under baryon number ~\cite{McKeen:2020oyr,McKeen:2021jbh,Goldman:2022brt}, or through DM capture from a clumpy halo with large boost factors~\cite{Bramante:2021dyx}, or due to other internal heating mechanisms independent of DM~\cite{2010A&Agonzalesheat,Hamaguchi:2019oev}. Observation of such NS would not only shed light on late evolutionary stages, but also on their equation of state. \newline

\section*{Acknowledgements}
We thank Fayez Abu-Ajamieh, Nirmal Raj and Peter Tinyakov for comments on the manuscript. S.C. and S.K.V. thank SERB Grant CRG/2021/007170 ``Tiny Effects from Heavy New Physics" from Department of Science and Technology, Government of India. 
R.G. is supported by MIUR grant PRIN 2017FMJFMW and
acknowledges partial support from the Spanish MCIN/AEI/10.13039/501100011033 grant PID2020-113334GB-I00. 
R.K.J. is supported by the Core Research Grant~CRG/2018/002200 and the Infosys Foundation, Bengaluru, India through the Infosys Young Investigator award. S.K.V. also acknowledges IoE funds from Indian Institute of Science. M.S.N.K. acknowledges the support from FCT - Fundação para a Ciência e a Tecnologia through Investigador contracts and exploratory project (IF/00956/2015/CP1273/CT0002).
R.G. thanks the Galileo Galilei Institute for hospitality during this work.

\bibliographystyle{apsrev4-1}
\bibliography{refs}

\clearpage
\appendix
\onecolumngrid

\twocolumngrid
\setcounter{equation}{0}
\setcounter{figure}{0}
\setcounter{table}{0}
\setcounter{section}{0}
\setcounter{page}{1}
\makeatletter
\renewcommand{\theequation}{S\arabic{equation}}
\renewcommand{\thefigure}{S\arabic{figure}}
\renewcommand{\thetable}{S\arabic{table}}
\renewcommand{\thesection}{S\arabic{section}}
\onecolumngrid

The appendix is organized as follows. In Sec.~\ref{sec:compip}, we describe computational inputs required to robustly estimate neutron star (NS) temperature due to dark matter (DM) heating. In Sec.~\ref{sec:bkg}, we detail background modelling and their impact on the signal-to-noise ratio (SNR).

\setcounter{equation}{0}
\setcounter{figure}{0}
\setcounter{table}{0}
\setcounter{section}{0}
\setcounter{page}{1}
\makeatletter

\section{Computational inputs}
\label{sec:compip}
\paragraph*{{\it 1. DM parameters $\rho_\chi$ and $v_d$:}} Study of local stellar kinematics provides one of the leading constraints on local density of DM and their velocity dispersion~\cite{Pato:2015dua,Buch:2018qdr}. 
Assuming spherically symmetric generalized NFW profile of the form $\rho_{\rm DM}\propto (r/r_s)^{-\gamma^\prime}(1+r/r_s)^{-3+\gamma^\prime}$, with $\gamma^\prime$ being the inner slope and $r_s$ the scale radius, dynamical constraints on the DM profile are placed by fitting the expected rotation curves, and surface density (at the Solar position) to observations~\cite{Pato:2015dua}. For fixed values of $r_s=20\,{\rm kpc}$, $\gamma^\prime=0.75$, we consider values for $\rho_\chi=0.39-0.52{\rm\,GeV/cm}^3$, with a flat prior. 
This range of values correspond to variations in the galactic parameters, namely the Solar distance from the galactic center  that ranges from $7.98$\,kpc to $8.68$\,kpc, and the local circular velocities $v_0= 214 -258 $\,km/s, respectively, for a representative baryonic model. Assuming Maxwell-Boltzmann velocity distribution for DM in the galactic rest frame, the dispersion velocity is obtained through the relation $v_d=\sqrt{3/2}\,v_0$.

\begin{figure}[h]
\begin{center}
\hspace{-0.8cm} \includegraphics[scale=0.38]{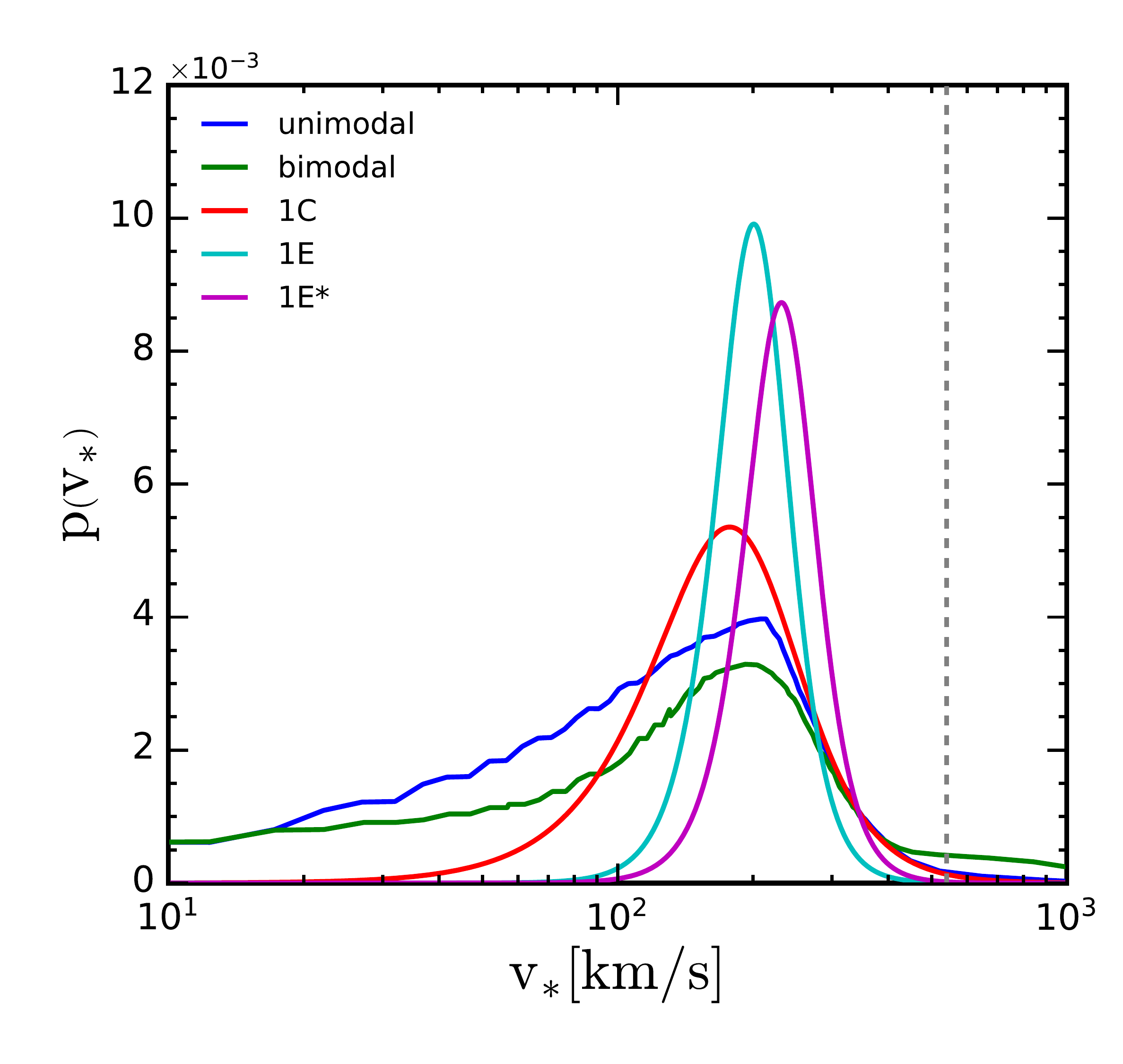} \includegraphics[scale=0.38]{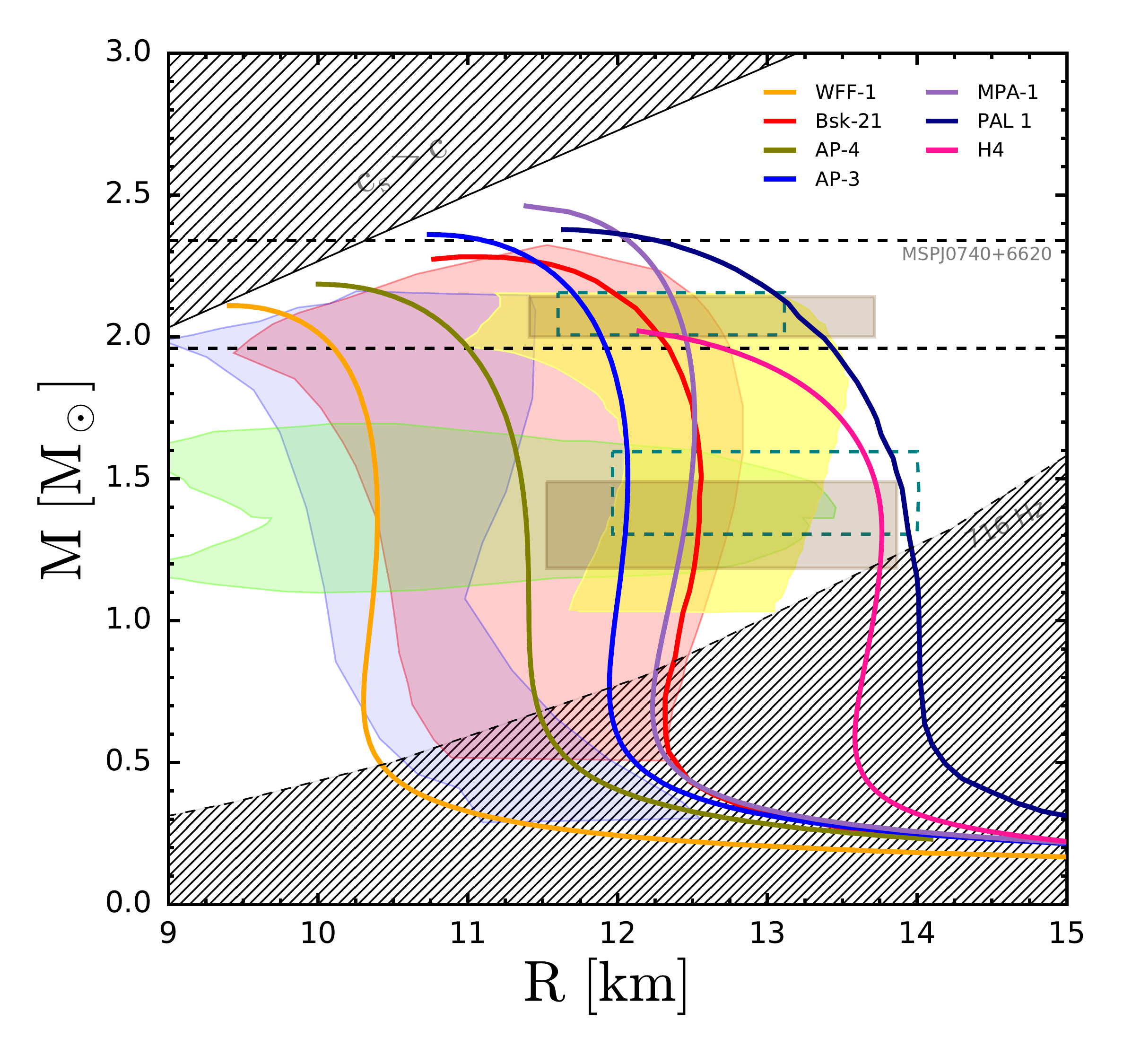} 
    \caption{{\it Left panel:} Normalized PDFs for velocities of old NS in the local bubble at the current epoch~\cite{Ofek:2009wt,2010A&A...510A..23S}. The dashed vertical line marks the Galactic escape velocity, taken here to be $540\,$km/s. {\it Right panel:} Mass-radius relation for representative EoS for nuclear matter at high densities, reproduced from~\cite{Ozel:2016oaf}. The blue shaded area indicates preferred region from observation of pulsars~\cite{Ozel:2016oaf}. The yellow region is preferred by the observation of binary NS merger where the fit was performed for purely hadronic EoS~\cite{Most:2018hfd}. The green region indicates the 90\% CL obtained from EoS insensitive fit from GW170817~\cite{LIGOScientific:2018cki}. 
    Bayesian fits at 90\% CL from combination of gravitational wave data and low energy nuclear and astrophysical data are shown as red regions~~\cite{Raithel:2018ncd}.
    The horizontal thick-dashed lines show the mass measurement  of the heaviest observed pulsar  MSP J0740+6620 of 2.14$^{+0.2}_{-0.18}$ M$_\odot$  at 95\% CL~\cite{NANOGrav:2019jur}. The dashed curve labelled 716 Hz represents the limit on the M-R of the fastest spinning pulsar obtained using observations from the Green Bank Telescope~\cite{hessels2006radio}. }
    \label{fig:EoS_M_R_all}
    \end{center}
\end{figure}

\paragraph*{{\it 2. NS velocity distributions $v_\star$:}} Population synthesis studies of galactic NS have been performed in several works which simulate the spatial and velocity distributions of old NS in the Milky Way~\cite{blaes1993ApJ,Ofek:2009wt,2010A&A...510A..23S,Tani2016RAA}.
Historically, these studies were motivated by the observational hunt for isolated, accreting NS in near X-ray bands~\cite{blaes1993ApJ,Treves:1999ne,Zane:1995bx}. From the simulations above, we obtain several normalized probability distribution functions (PDFs) for NS velocities, at the current epoch. We choose two PDFs obtained in ref.~\cite{Ofek:2009wt}, where the unknown initial NS velocities were modeled as unimodal and bimodal distributions. Similar studies of NS velocity distributions using different Galactic potential, progenitor distributions, and birth velocities, were reported in~\cite{2010A&A...510A..23S}. From those, we pick three models, namely 1C, 1E, and 1E*, which are representative of the full range of models studied in~\cite{2010A&A...510A..23S}. The PDF for model 1C corresponds to progenitor velocity distribution F06E~\cite{Faucher-Giguere:2005dxp} and galactic potential P90~\cite{P90}. While the PDF for models 1E and 1E* correspond to F06P~\cite{Faucher-Giguere:2005dxp} and P90. The models 1E* and 1E differ by 20\% in their choice of escape of velocity at the Solar position and by 80\% for the virial mass of the milky way, respectively. We present these PDFs in the left panel of fig.~\ref{fig:EoS_M_R_all}.\\

\paragraph*{{\it 3. NS Mass-Radius relationships:}} 
In the right panel of fig.~\ref{fig:EoS_M_R_all}, we show the mass-radius relationship for NS composed predominantly of neutrons ($\sim 90\%$) and proton+lepton ($\sim 10\%$) matter. The multi-colored curves in this figure correspond to different assumptions of the physics and computational techniques, and are compiled in~\cite{Ozel:2016oaf,Lattimer_2001}. The EoS WFF-1 ~\cite{wiringa1988equation} (orange) is obtaine   by variational methods.  EoS BSk-21 (red) is phenomenologically derived through fits to several low energy nuclear data~\cite{Potekhin:2013qqa,Potekhin:2015qsa}. The AP4 EoS (olive green) is akin to WFF-1, simultaneously applying many-body and relativistic corrections~\cite{akmal1997spin}. EoS AP-3 (blue solid), is similar to AP-4, but it employs a different central pressure, energy and baryon density. The EoS MPA-1 (purple) is derived from microscopic ab-initio calculations based on Dirac-Brueckner-Hartree-Fock methods that include relativistic effects~\cite{muther1987nuclear}.
The phenomenological non-relativistic EoS  PAL-1~\cite{prakash1988equation} is shown in navy blue. The EoS H4 (pink) is obtained by adding hyperon matter to the canonical neutron and proton matter~\cite{Lackey:2005tk}. 

While the multi-colored curves represent theoretical considerations, it is necessary to understand how much of it is supported by observations. The shaded regions display the M-R values supported by various observations and experiments. The green shaded area show 90\%  confidence limits obtained by performing a fit for the tidal deformability and mass using EoS-insensitive methods from recent gravitational wave observations of mergers of binary NS~\cite{LIGOScientific:2018cki,De:2018uhw}. Tidal deformability data fit to purely hadronic EoS using the latest outer core model~\cite{Drischler:2017wtt} results in yellow shaded regions at 3-$\sigma$~\cite{Most:2018hfd}.
Bayesian fits obtained by combining gravitational wave data with low energy nuclear and astrophysical data are shown as red regions~\cite{Raithel:2018ncd} at 90\% confidence level. Radio and X-ray observation of pulsars~\cite{Ozel:2016oaf} are shown by light blue area. Simultaneous mass-radius measurements of PSR  J0030+0451 and J0740+6620 by NICER~\cite{miller2019psr,miller2021radius} (\cite{riley2019nicer, riley2021nicer}) are shown as green dashed and brown shaded rectangular regions representing 68\% confidence limits, respectively. \\

\section{Background modelling}\label{sec:bkg}

Given any sky co-ordinates the JWST background tools provide us with equivalence in-field radiance of background emission as a function of wavelength for every day of the year \textit{(https://jwst-docs.stsci.edu/jwst-general-support/jwst-background-model)}. For wavelengths $\sim 1\, \mu$m the emissions are dominated by Zodiacal light. We have chosen a well studied blank field that lies close to Fornax. 
\begin{figure*}[h]
    \centering
    \includegraphics[scale=0.33]{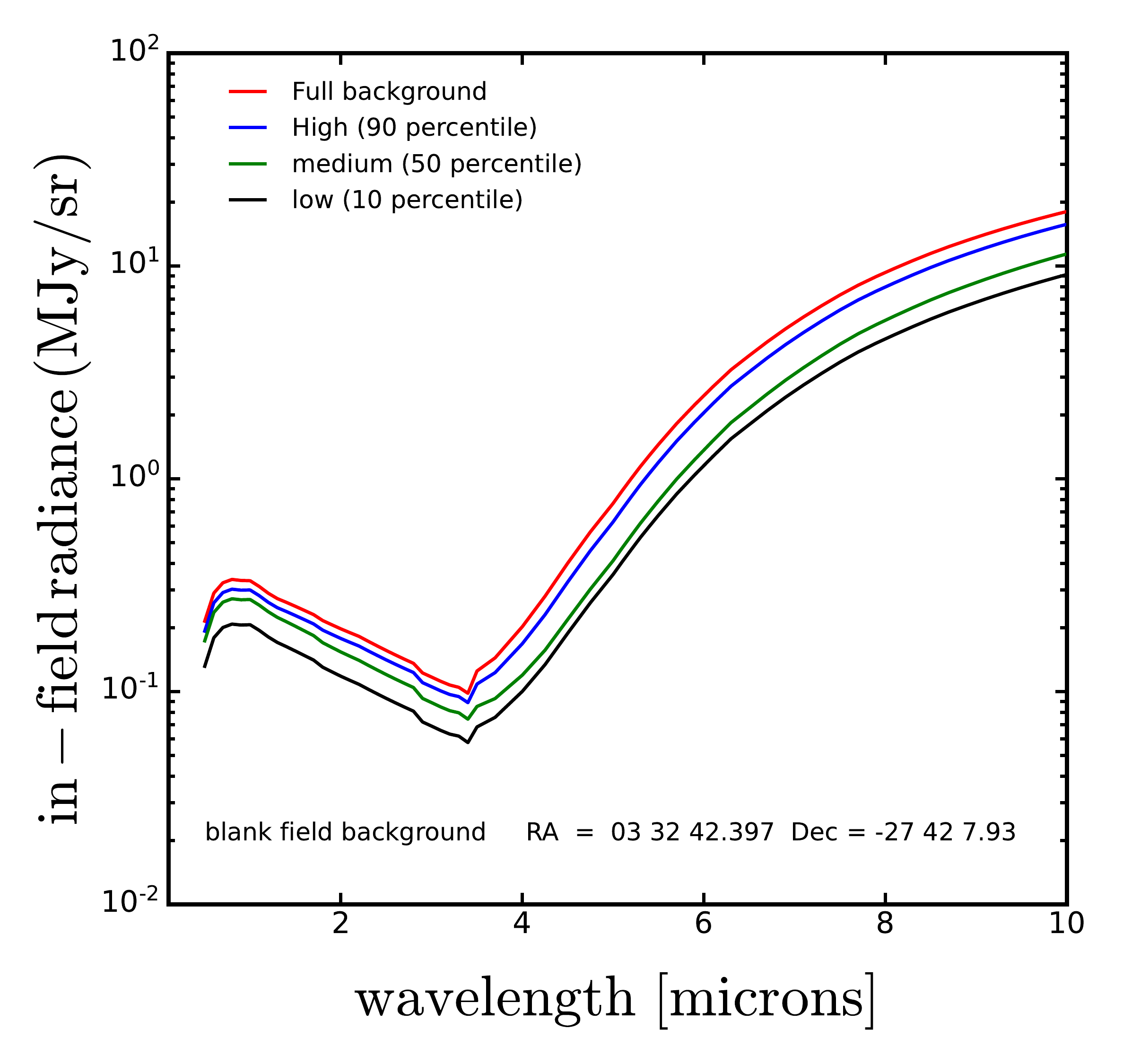}  \includegraphics[scale=0.33]{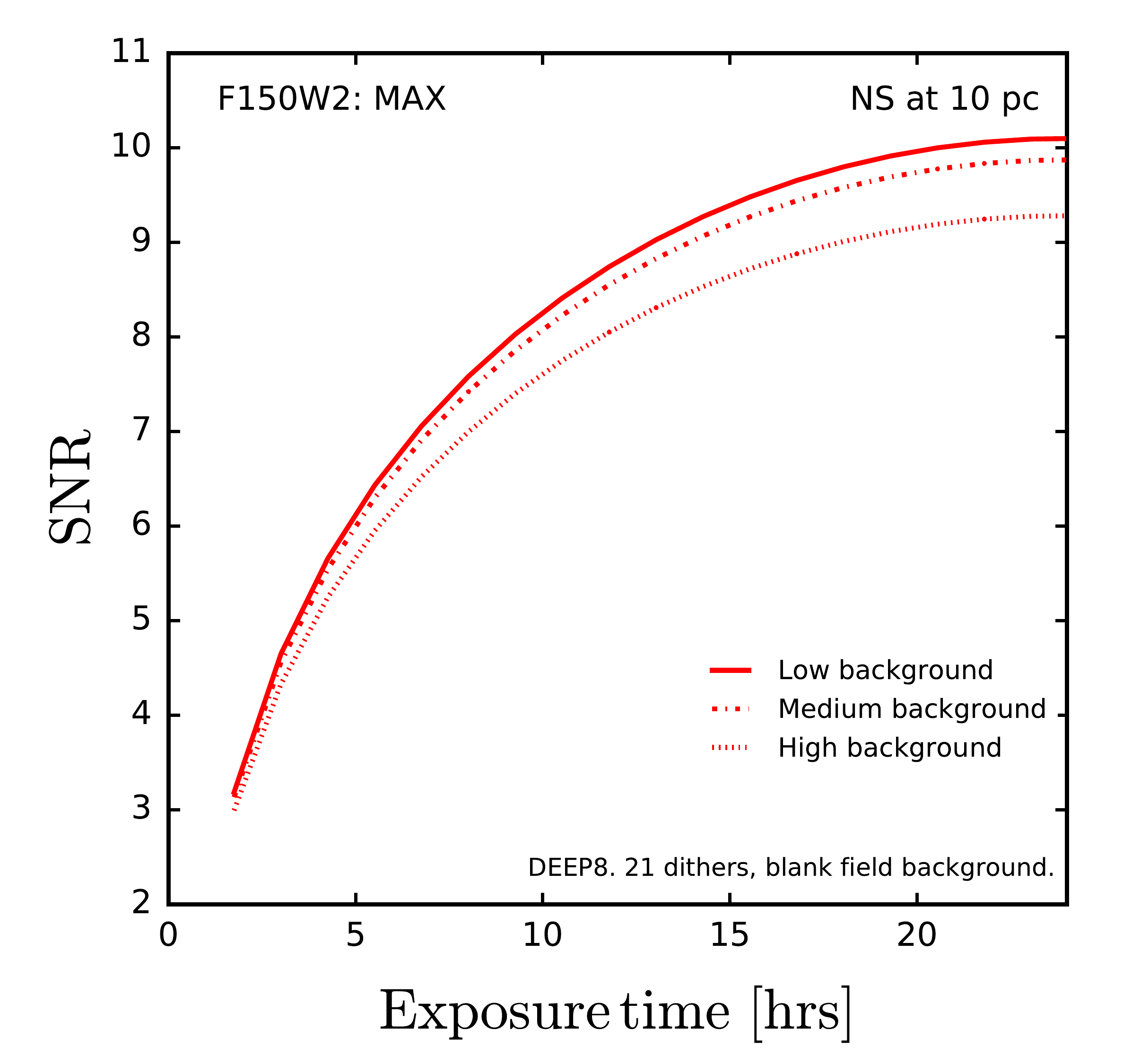}
    \caption{Background as function of wavelength (left). See text for details. Dependence of SNR on background percentile (right).}
    \label{fig:my_bkg}
\end{figure*}

For a given wavelength, we first construct a histogram of background emission over all feasible days of observations. Since we do not know the exact days of possible observations, we statistically classify three background models that delimits annual variations in background emissions. The background model `high' corresponds to considering only 90$^{\rm th}$  percentile of the above distribution. Similarly, the models `medium' and `low' correspond to 50$^{\rm th}$ and 10$^{\rm th}$ percentile, respectively. We show this in the left panel of Fig.~\ref{fig:my_bkg}.

In the right panel of Fig.~\ref{fig:my_bkg}, we show SNRs for the MAX scenario corresponding to low (solid), medium (dashed), and high (dotted) background models. As mentioned in the main text the maximum attainable SNR within a day of observation varies by at most 2 units.

\end{document}